\documentstyle[twocolumn,aps,epsfig,array]{revtex}

\newcommand{\be}{\begin{equation}}
\newcommand{\ee}{\end{equation}}
\newcommand{\ba}{\begin{eqnarray}}
\newcommand{\ea}{\end{eqnarray}}

\draft

\title{Thermodynamical Test of Non Extensive Thermostatistics}
\author{S. Mart\'{\i}nez${^1}$, F.
Pennini${^1}$, A. Plastino${^1}$ and H.
Vucetich${^2}$}
\address{$^1$ La Plata Physics Institute, La Plata National University and
CONICET\thanks{Argentine National Research Council},\\ C.C. 727,
1900 La Plata, Argentina. \\ $^2$ Astronomical Observatory, La
Plata National University\\ Paseo del Bosque S/N, C.P. 1900, La
Plata,  Argentina. }

\begin{document}
\maketitle

\begin{abstract}
An ideal mixture of parahydrogen (with nuclear spin $K=0$) and
orthohydrogen (with $K=1$), in statistical weights $1/4$ and
$3/4$, respectively, is used as a test ground for the existence of
non-extensivity in chemical physics. We report on a new bound on
the non extensivity parameter $q - 1$ that characterizes
generalized thermostatistics \`a la Tsallis. This bound is
obtained on the basis of laboratory measurements of the specific
heat of hydrogen. Suggestions are advanced for the performance of
improved measurements.


\end{abstract}
\vskip 3mm

    In the last twelve years a considerable amount of
effort has been expended investigating systems for which the
traditional Boltzmann-Gibbs-Shannon (BGS) statistical treatment,
for a variety of circumstances, seems to fail, al least in some
respects \cite{PP99}. The so-called Tsallis non-extensive
thermostatistics (TT) has played a prominent role in these
endeavors \cite{T99}. TT is characterized by a real parameter $q$,
such that $q=1$ yields the standard BGS treatment. The first
successful applications of such formalism was the finding of a
non-divergent, physical distribution function for stellar
polytropes \cite{PP93}. A significant TT step was that of
Boghosian. His analysis of the relaxation of a two dimensional
pure electron plasma \cite{t8} yielded a $q$-value quite different
from unity.

 Many subsequent advances, both on the basic theory front and in
different applications, have transformed this non-extensive
treatment into a new paradigm for statistical mechanics
\cite{PP99,T99,t1}. TT is based upon Jaynes' MaxEnt approach to
statistical mechanics \cite{K67}. Its main feature is that of
replacing the conventional Shannon logarithmic information measure
$S$ (subject to appropriate constraints derived from the available
a priori knowledge) by Tsallis' non-extensive measure (we
normalize all quantities to $k=1$)
\begin{equation}
S_q=(q-1)^{-1}\, \sum_{i=1}^W (p_i\,-\,p_i^q),\,\,q \in {\Re}
\label{Sq},
\end{equation}
where the set $\{ p_i \}$, to be obtained by recourse to Jaynes'
variational procedure, yields the probability distribution
associated to the relevant $W$ microstates. Notice that for $q=1$
one has $S_q \equiv S$.

$S_q$ is non-extensive on account of the fact that, for a system
composed of two noninteracting subsystems $(A,B)$ \cite{PP99}
\begin{equation}
S_q(A + B) = S_q(A) + S_q(B) + (1 - q)  S_q(A) S_q(B),
\label{Tsall-S}
\end{equation}
so that $q$ measures the degree on non-extensivity. During almost
10 years, the concomitant non extensive thermostatistical
considerations were based upon the theoretical framework of Curado
and Tsallis \cite{t3}. Nowadays, it is believed that this
framework has been superseded by a new one, called ``the
normalized one'', advanced in \cite {pennini}, and formalized in
great detail by Tsallis, Mendes and Plastino \cite{mendes}, which
seems to exhibit important advantages \cite {PTV01}. This
normalized treatment, in turn, has been considerably improved by
the so-called ``Optimal Lagrange Multipliers'' (OLM) approach
\cite{OLM}. The OLM treatment seems to provide one with a natural
bridge that connects Tsallis' thermostatistics with {\it
thermodynamics} \cite {OLM,equip,gasi,ley0}. The main difference
between the Curado-Tsallis approach \cite{t3} and the
``normalized-OLM'' \cite{OLM} one is to be found in the manner of
computing expectation values \cite{OLM}. One has, for a given
observable $A$ that takes the value $a_i$ for the microstate $i$
\begin{equation}
\langle A \rangle_q\,=\,\sum_{i=1}^W\,p_i^q
\,a_i,\,\,\,\mbox{(Curado-Tsallis)} \label{gev},
\end{equation}
and
\begin{equation}
\langle A \rangle_q\,=\,\frac{\sum_{i=1}^W\,p_i^q
\,a_i}{\sum_{i=1}^W\,p_i^q},\,\,\,\mbox{(normalized-OLM)}
\label{gevOLM}.
\end{equation}

Differences in some details of the variational procedure
distinguish the approach of Ref. \cite{mendes} from the one of
\cite{OLM}, that is to be the focus of attention in this work.

 It is quite clear that $S_q$ provides us with an exceedingly useful
information theory tool. Can we identify it with a thermodynamic
entropy as well? For an appropriate answer an experimental
determination of the $q$-value within a purely thermodynamics setting
becomes necessary.  One such a test was reported in
Refs. \cite{PPV95,TSBL95}, where bounds to $|q-1|$ were established
using the cosmic black-body radiation. Also, present-day
determinations of the Stefan-Boltzmann constant $\sigma$ put a
similar constraint, for which the order of magnitude is $|q-1|
<10^{-4}$. Finally, bounds on the non extensivity of fermion systems
have been reported \cite{PTV97}, namely \(|q - 1| <
2\times10^{-5}\). All these works refer to the Curado-Tsallis
formalism. With respect to this treatment, the OLM reformulation of
the theory \cite{OLM} entails a sensible {\it physical difference}: in
the latter, non extensivity is restricted just to $S_q$, while the
energy behaves as an extensive quantity (it is non-extensive in the
Curado-Tsallis instance). To our knowledge, no comparison of
predictions arising from this OLM scenario with experiment has been
carried out yet.

    Now, except for the measurement of Stefan-Boltzmann constant, all
the above reported bounds on \(q-1\) have been based on {\it
observational} rather than on {\it experimental}, chemical physics
data. The advantage of laboratory measurements in the comparison
of a new theory with nature lies in the possibility of varying and
controlling the conditions in which the data is acquired, thus
ensuring a thorough validity test. With regards to the
observational tests, mostly of cosmological nature
\cite{PPV95,TSBL95,PTV97,TV98}, it has been forcefully argued
\cite{HaBa96} that  they  can {\it not} test the degree of non
extensivity  of a system, as the application of thermodynamic
arguments is of a strictly local character. This entails that
limits on the non-extensivity, in a large, cosmological scale, can
not be thereby obtained.

    In order to test non extensivity, a (laboratory) physical system
composed of two different weakly interacting subsystems should be
used.  It is well known that H$_2$ gas is a mixture of two different
substances \cite{Pathria,Huang}: parahydrogen (with nuclear spin
$K=0$) and orthohydrogen (with $K=1$), in statistical weights $1/4$
and $3/4$, respectively. Since there exist accurate (about 1\%)
measurements of both its specific heat and relevant spectroscopic
data, this ideal mixture is a very good test ground for the existence
non-extensivity in chemical physics. In this communication a new bound
on the non extensivity parameter $q - 1$, based on laboratory
measurements of the specific heat of hydrogen, is reported, and
suggestions are advanced for the performance of improved measurements.

    Since the proposal of Giauque \cite{GiauqueFE}, it has been
customary to compute the thermodynamic properties of simple
substances, such as hydrogen, from spectroscopic data. Indeed, when
accurate spectroscopic data are available, this method gives much
greater precision than direct thermal measurements. On the other hand,
the method rests on the validity of BGS statistics as the correct
description of nature. Giauque himself computed the thermodynamic
properties of the H$_2$ molecule \cite{GiauqueH2} and found very good
agreement with the measured data
\cite{Eucken12,EuckHill,CornEast}. The Giauque proposal has been very
successful, to the point that no new direct measurement of the
specific heat of H$_2$ molecule has been reported since 1930!

    At this point we start our non-extensive considerations. Since the
experimental thermodynamic value of $q$ can not be expected to
appreciably differ from unity (otherwise, non-extensivity would have
been discovered decades ago!), a perturbative treatment around $q=1$
will be undertaken.  In order to compute the specific heat of a
substance in the standard Curado-Tsallis (CT) formalism, the point of
departure is given by the Tsallis partition function
\begin{equation}
Z^{(CT)}_q = \sum_j e_q^{- \beta \epsilon_j}, \label{Def-Z(2)}
\end{equation}
where $e_q^x \equiv \left[1 + (1-q)x \right]^{1/(1-q)}$,
$\beta=1/T$ with $T$ the temperature and $\{\epsilon_j\}$ is the
set of energy levels. We compute then the Tsallis free energy
\begin{equation}
F^{(CT)}_q = -\frac{1}{\beta} \ln_q Z^{(CT)}_q, \label{Def-F(2)}
\end{equation}
where $\ln_q x \equiv (x^{1-q} - 1)/(1 - q)$, and finally, apply
the usual thermodynamic relations
\begin{equation}
S^{(CT)}_q = - \frac{\partial F_q^{(CT)}}{\partial T}, \qquad
C^{(CT)}_q = T \frac{\partial S_q^{(CT)}}{\partial T}.
\end{equation}

An straightforward expansion of equations (\ref{Def-Z(2)}) and
(\ref{Def-F(2)}) around $q=1$ yields
\begin{equation}
F^{(CT)}_q = F - \frac{1}{2} (q - 1) \left[\frac{C}{\beta} + \beta
\left(U^2 - F^2\right) \right].
                    \label{Pert-F(2)}
\end{equation}
where $U$ and $F$ are the BGS $(q=1)$ internal and free energies
respectively and $C$ is the BGS specific heat.

Additionally, for an out of equilibrium $(f, 1-f)$ mixture of two
substances $(A,B)$, the specific heat satisfies the analog of
equation (\ref{Tsall-S})
\ba
&& C^{(CT)}_q (A + B)= C^{(CT)}_q (A) + C^{(CT)}_q (B) + (1 - q) {\cal C}_{AB},
\label{C(2)-Mix} \\
&&\label{Cmix2} {\cal C}_{AB} =
\left[C^{(CT)}_q (A) S^{(CT)}_q (B) + C^{(CT)}_q (B) S^{(CT)}_q
(A) \right].
\ea

    The calculation of the specific heat according to the OLM treatment
    is of a somewhat more
 subtle nature because the thermodynamic relations does not keep the usual
 form in terms of Tsallis' entropy \cite{OLM}.
 One starts from the definition
\begin{equation}
C_q=\frac{\partial U_q}{\partial T}=-\beta^2 \frac{\partial
U_q}{\partial \beta},
\end{equation}
where the internal energy $U_q$ can be written as \cite{OLM}
\begin{equation}
U_q=\frac{\sum_j f_j^{q/(1-q)} \epsilon_j}{\sum_j f_j^{q/(1-q)}}.
\end{equation}

 In this equation, the  $f_j= 1-(1-q)\beta \Delta_j$ are the so-called configurational characteristics
\cite{OLM} where $\Delta_j=\epsilon_j-U_q.$
Taking the pertinent derivatives, one arrives to
\begin{equation}
\frac{C_q}{q \beta^2}=\frac{\left\langle \Delta_j^2
f_q^{-1}\right\rangle_q}{1-q
\beta \left\langle \Delta_j f_q^{-1}\right\rangle_q}.
\label{Cq}
\end{equation}
which in the $q \rightarrow 1$ limit goes into the BGS specific heat.
No corrections for the fact that we deal with a mixture are necessary,
since the internal energy is additive in the OLM approach \cite{OLM}
\begin{equation}
U_q^{\rm OLM} (A + B) = U_q^{\rm OLM} (A) + U_q^{\rm OLM} (B).
                    \label{Adit-U-OLM}
\end{equation}

    The specific heat of parahydrogen (orthohydrogen)
can now be written in the fashion
\begin{equation}
C_q = C + (q - 1) \Delta C,
\end{equation}
with
\begin{eqnarray}
\Delta C^{(CT)} &=& 2\beta\left(F C + U \beta \frac{d
    C}{d\beta}\right)\nonumber\\
& &  - \left[{C^2} + \beta \frac{d C}{d\beta} - \beta
    \frac{d}{\beta} \left(\beta \frac{d C}{d\beta}\right)
    \right],
                \label{DeltaCv2}
\end{eqnarray}
for the Curado-Tsallis version  of the formalism, and, with
$\Delta^e_j = (\epsilon_j-U)$,
\begin{eqnarray}
\Delta C^{\rm OLM} &=& C - 2 \beta^3 \left\langle {\Delta^e_j}^3
    \right\rangle + \nonumber\\
& & \beta^4 \left(\frac{1}{2} \left\langle
    {\Delta^e_j}^4 \right\rangle - \frac{3}{2} \left\langle {\Delta^e_j}^2
    \right\rangle^2\right), \label{DeltaCv3}
\end{eqnarray}
for its OLM counterpart.  Notice that all  quantities entering
$\Delta C$ are evaluated within the framework of the standard,
BGS statistics. For the case of H$_2$ one has now
\begin{eqnarray}
\Delta C^{(CT)} ({\rm H}_2) &=& \frac{1}{4}\Delta C^{(CT)}_{\rm p}
+ \frac{3}{4}\Delta C^{(CT)}_{\rm o}\nonumber\\
& & - \frac{3}{16} \left(C_{\rm p} S_{\rm o} + C_{\rm o} S_{\rm p} \right),\\
\Delta C^{\rm OLM} ({\rm H}_2) &=&  \frac{1}{4}\Delta C^{\rm OLM}_{\rm p}
+ \frac{3}{4}\Delta C^{\rm OLM}_{\rm o},
\end{eqnarray}
where $C_{\rm o},C_{\rm p}$ ($S_{\rm o},S_{\rm p}$) are the BGS specific
heats (entropies)
of ortho- and parahydrogen, respectively.

    The structure of $\Delta C$ in Eq. (\ref{DeltaCv2}) exhibits in
a quite clear fashion i) a characteristic non-extensive behavior and
ii) a dependence on the reference point for the energy. In
Eq. (\ref{DeltaCv3}), on the other hand, the non-extensivity is not
apparent! It is hidden in the higher order momenta of the
energy. Also, the troublesome dependence on the origin of the energy
scale just vanishes.

\begin{figure}
\epsfig{file=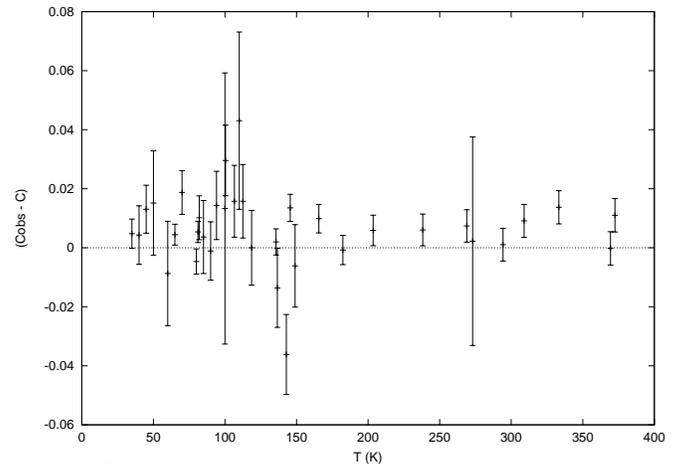,scale=.7}
\caption{O-C residuals of the specific
heat of H$_2$, with respect to the Bolztmann-Gibbs-Shannon theory.}
\label{H2Graph}
\end{figure}

    Table \ref{exp-data} displays the accurate data used in this
work. The theoretical BGS thermodynamic functions were computed
from spectroscopic data using the usual expressions. The constants for
the rotational levels were taken from Ref.  \cite{HerzbergSDM}, since
they are accurate enough to get a good representation of experimental
data. The O-C differences between BGS theory and experiment
(Fig. \ref{H2Graph}) were used to fit the Tsallis parameter $(q-1)$
with a simple least squares procedure. The ensuing results are shown
in the first line of table \ref{H2Results}.

    These results, however, may be contaminated by systematic
errors. The largest deviations from the BGS specific heat
come from the data in reference \cite{CornEast}. Assuming that in
this data there is a small deviation from the $3/4 - 1/4$ mixture,
the O - C differences would adopt the appearance
\begin{equation}
\Delta C_{obs} = (q-1) \Delta C + \epsilon (C_{\rm p} - C_{\rm
o}),
\end{equation}
where $\epsilon$ is an auxiliary parameter. In such a case one
obtains the results of lines 2 and 3 of table \ref{H2Results},
which are consistent with zero at the 95\% CL.

This suggest the existence of  systematic errors in the data.

    As a final result, we can state that the Tsallis parameter is
consistent with zero at the 95\% confidence limit:
\begin{eqnarray}
\mid{q-1}\mid_{CT} &<&  5 \times 10^{-4} \qquad  (95\% {\rm C.L.})
                        \label{Lim:q-1:2} \\
\mid{q-1}\mid_{OLM}& <& 1.2 \times 10^{-3} \quad    (95\% {\rm C.L.}).
                        \label{Lim:q-1:3}
\end{eqnarray}
The above  equations constitute the main result of this
communication.

    Our results show that the BGS statistics is,
among the set of theories parameterized by $q - 1$, the one that
describes the available data with an accuracy better than
$10^{-3}$. Within this level of accuracy, there is no evidence on a
non-extensive behavior in this simple system.

    With better experiments, the accuracy of our results
(\ref{Lim:q-1:2}) and (\ref{Lim:q-1:3}) could be improved, in
principle, by several orders of magnitude. Indeed, spectroscopic data
have a potential accuracy of 10$^{-6}$, while an accuracy of 10$^{-4}$
should be easily attained with the use of modern experimental
techniques. In such a way a stricter test for the validity of the
BGS statistics (or, alternatively, for the existence of
non-extensivity) would become available for this system.

\begin{table}
\begin{tabular}{ll@{$\pm$}lrll@{$\pm$}lrll@{$\pm$}lr}
$T$ (K) & \multicolumn{2}{c}{$C$} & Ref & $T$ (K) &
\multicolumn{2}{c}{$C$} & Ref & $T$ (K) & \multicolumn{2}{c}{$C$}
& Ref \\ \hline
  35.00 & 1.505 & 0.005 &\cite{Eucken12}&   94.12 & 1.688 & 0.012
&\cite{EuckHill}&  165.58 & 2.147 & 0.005 &\cite{CornEast}\\
  40.00 & 1.505 & 0.010 &\cite{Eucken12}&  100.00 & 1.727 & 0.046
&\cite{Eucken12}& 182.41 & 2.213 & 0.005 &\cite{CornEast}\\
  45.00 & 1.515 & 0.008 &\cite{Eucken12}&  100.42 & 1.746 & 0.012
&\cite{EuckHill} &  203.63 & 2.298 & 0.005 &\cite{CornEast}\\
  50.00 & 1.520 & 0.018 &\cite{Eucken12}&  106.64 & 1.777 & 0.012
&\cite{EuckHill} &  238.23 & 2.390 & 0.005 &\cite{CornEast}\\
  60.00 & 1.510 & 0.018 &\cite{Eucken12} &  110.00 & 1.828 & 0.030
&\cite{Eucken12} &  269.02 & 2.444 & 0.005 &\cite{CornEast}\\
  65.00 & 1.535 & 0.004 &\cite{Eucken12}&  112.52 & 1.819 & 0.012
&\cite{EuckHill} &  273.10 & 2.444 & 0.035 &\cite{Eucken12}\\
  70.00 & 1.565 & 0.007 &\cite{Eucken12}&  118.63 & 1.847 & 0.004
&\cite{EuckHill} &  294.27 & 2.466 & 0.005 &\cite{CornEast}\\
  80.00 & 1.586 & 0.004 &\cite{Eucken12}&  135.71 & 1.967 & 0.004
&\cite{CornEast} &  308.96 & 2.486 & 0.006 &\cite{CornEast}\\
  81.12 & 1.601 & 0.004 & \cite{CornEast}&  136.62 & 1.957 & 0.013
&\cite{EuckHill} &  333.31 & 2.505 & 0.006 &\cite{CornEast}\\
  82.00 & 1.611 & 0.007 &\cite{Eucken12}&  142.89 & 1.975 & 0.014
&\cite{EuckHill}  &  369.40 & 2.504 & 0.006 &\cite{CornEast} \\
  85.00 & 1.621 & 0.012 &\cite{Eucken12} &  145.64 & 2.041 & 0.005
&\cite{CornEast} &   372.52 & 2.516 & 0.006 &\cite{CornEast}\\
  90.00 &  1.646 & 0.010 & \cite{Eucken12}&  148.98 & 2.041 & 0.014
&\cite{EuckHill}
\end{tabular}

\caption{Experimental data on the H$_2$ specific heat. The columns
display, respectively, the temperature in K, the molar specific
heat in units of $R$ with our adopted standard deviation, and the
corresponding bibliographic reference.} \label{exp-data}
\end{table}

\begin{table}
\begin{center}
\begin{tabular}{lcc}
Parameter & CT Version & OLM Version\\
\hline
$q-1$  & $-0.21 \pm 0.05$  & $0.89 \pm 0.23$\\
\hline
$q-1$      & $-0.11 \pm 0.07$  & $0.52 \pm 0.57$ \\
$\epsilon$  & $1.0 \pm 3.7$  & $2.9 \pm 4.1$\\
\hline
\end{tabular}
\end{center}
\caption{Results of the analysis performed in this work. The
columns of the table display the adjusted parameters' names and
their values (in units of 10$^{-3}$) for the two versions, CT and
OLM, of the theory. The first line lists a raw fit to the
residuals, while the second and third lines display a fit that
assumes a small departure from the $1/4 - 3/4$ proportions of the
mixture.} \label{H2Results}
\end{table}

\end{document}